\documentclass[a4paper,10pt,twoside]{cpc-hepnp}

\usepackage{multicol}
\usepackage{graphicx}
\usepackage{booktabs}
\usepackage{amssymb,bm,mathrsfs,bbm,amscd}
\usepackage[tbtags]{amsmath}
\usepackage{lastpage}
\usepackage{CJK}

\begin{document}

\fancyhead[c]{\small Submitted to Chinese Physics C} \fancyfoot[C]{\small 010201-\thepage}

\footnotetext[0]{Received 14 March 2009}

\title{the high-speed after pulse measurement system for PMT\thanks{Supported by National Natural Science Foundation of China (10775181) }}

\author{%
      CHENG Ya-Ping$^{1,2,3)}$
\quad QIAN Sen$^{1,2;1)}$\email{qians@ihep.ac.cn}%
\quad NING Zhe$^{1,2}$ \quad XIA Jing-Kai$^{1,2,3}$ \\
\quad WANG Wen-Wen$^{1,4}$ \quad WANG Yi-Fang$^{2}$
\quad CAO Jun$^{2}$
 \quad
JIANG Xiao-Shan$^{1,2}$  \\
\quad WANG Zheng$^{1,2}$ \quad LI Xiao-Nan$^{1,2}$
\quad QI Ming$^{4}$ \quad HENG Yue-Kun$^{1,2}$ \\
\quad ZHAO Tian-Chi$^{2}$ \quad LIU Shu-Lin$^{1,2}$
\quad LEI Xiang-Cui$^{1,2}$ \quad WU Zhi$^{1,2}$ }
\maketitle

\address{%
$^1$ State Key Laboratory of Particle Detection and Electronics, Beijing 100049, China\\
$^2$ Institute of High Energy Physics, Chinese Academy of Sciences, Beijing 100049, China\\
$^3$ University of Chinese Academy of Sciences, Beijing 100049,China\\
$^4$ Nanjing Unversity, Nanjing 210093, China\\
}

\begin{abstract}
A system employing a desktop FADC has been developed to investigate
the features of  8 inches Hamamatsu PMT. The system stands out for
its high-speed and informative results as a consequence of adopting
fast waveform sampling technology. Recording full waveforms allows
us to perform digital signal processing, pulse shape analysis, and
precision timing extraction. High precision after pulse time and
charge distribution characteristics are presented in this
manuscript. Other photomultipliers characteristics, such as dark
rate and transit time spread, can also be obtained by exploiting
waveform analysis using this system.
\end{abstract}



\footnotetext[0]{\hspace*{-3mm}\raisebox{0.3ex}{$\scriptstyle\copyright$}2013
Chinese Physical Society and the Institute of High Energy Physics
of the Chinese Academy of Sciences and the Institute
of Modern Physics of the Chinese Academy of Sciences and IOP Publishing Ltd}%

\begin{multicols}{2}

\section{Introduction}

As light sensors with outstanding characteristics such as high gain,
low noise and fast response, photomultiplier tubes (PMT) gain
popularity in nuclear and particle physics experiments, astronomy
and medical diagnostics. Taking advantage of PMTs' individual photon
detection capability, neutrino physics experiments\cite{lab21,lab1}
employ hundreds of photomultiplier tubes to detect low intensity
light, produced from large volumes of pure water, such as
Super-Kamioka Neutrino Detection Experiment or Liquid scintillators,
such as Daya Bay Reactor Neutrino Experiment.

Before PMTs start data taking mission, a series of tests have to be
made to check whether every single PMT have reached their desired
specifications, the gain, the transit time spread(TTS), the dark
count or dark current and so on\cite{lab19}. Time and charge
information from PMTs are used to reconstruct event energy and
vertex. After pulse occurs some time later after the initial
photoelectron signal, and cannot be distinguished from true physical
signals\cite{lab6}. Thus for low background neutrino experiments
that use large number of PMTs£¬after pulse is troublesome
background, and its features need fully studied and put into
simulation to evaluate its impacts\cite{lab7,lab8}. The mechanism of
after pulse was found to be the ionization of the residual gases by
the accelerated photoelectrons occurring inside the
PMT\cite{lab9,lab10}.

 In order to generate at least 20 $\mu$s time windows, the oscilloscopes
 were introduced even if their data taking speed was not so fast\cite{lab20}.In case of big batch PMT tests, high speed becomes prominent issue
for the test system. Compared with conventional oscilloscope test
system, the flash ADC test system can continue without dead-time.
While all oscilloscopes have dead-time between repetitive
acquisition of waveforms and their dead-times can sometimes be
orders of magnitude longer than acquisition time\cite{lab2}. In our
lab we have successfully implemented a high speed test system based
on a desktop flash ADC(FADC).

\section{Test Facility}
The test system was based on a 1 GHz desktop FADC (CAEN
DT5751)\cite{lab3}.At 1.0$\times10^7$ gain, the waveforms from the
anode of the PMTs was about 10 mV amplitude. With the digitizer's 1
Vpp input dynamics and a programmable DC-offset , signals as large
as 100 p.e. could be directly sampled by FADC and then read into the
DAQ computer and analyzed using Root.

 A laser diode was
used to light the PMT. A pulse generator (model AFG3102) was used to
drive the laser diode and its synchronizing signal after converting
to NIM level served as the trigger signal of FADC. The voltages to
the PMTs were supplied by the high voltages system SY1527 controlled
by the DAQ computer as well as the intensity and frequency of the
light source. With this coincidence, triggers caused by noise can be
largely suppressed. The pulse driving laser diode is 10 ns width
pulse. Its voltage was carefully tuned to get single photon and
multi photons.

\begin{center}
\includegraphics[width=7cm]{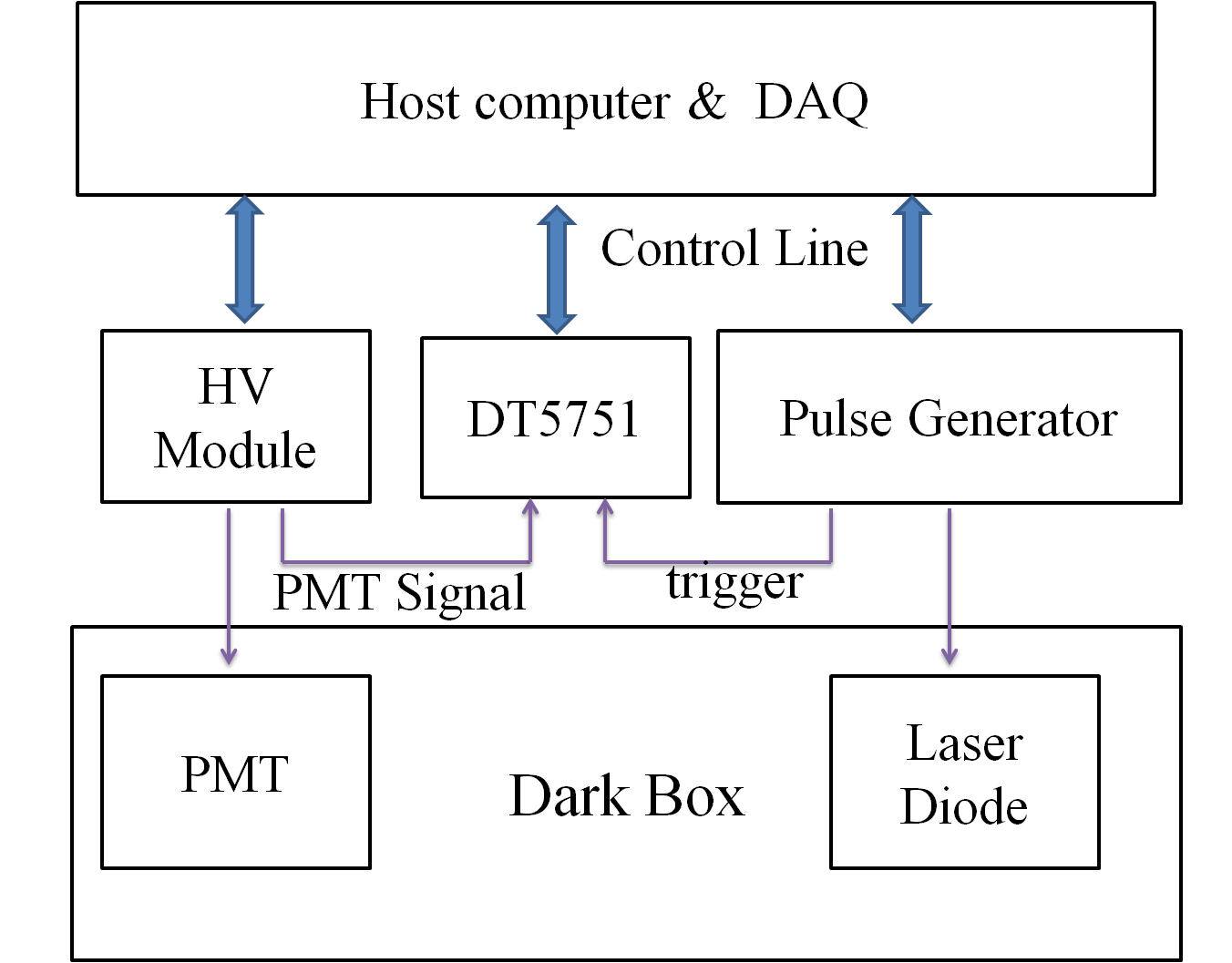}
\figcaption{\label{fig1}   The setup used for PMT performance
measurements. }
\end{center}

Another option for the test instrument was Lecroy WavePro 7100.
Their technical performances and specifications were listed in
Table~\ref{tab1}\cite{lab4,lab5}. Since we need speed up our test
procedure and demand for a large time buffer memory, DT5751 was our
final choice.
\\
\\
\\
\\
\\
\\

\begin{center}
\tabcaption{ \label{tab1}  Comparison of performance parameters for
DT5751 and WavePro 7100.} \footnotesize
\begin{tabular*}{80mm}{l@{\extracolsep{\fill}}rr}
\toprule \hphantom{0}Instrument & DT5751   & WavePro 7100   \\
\hline
Sampling Frequency & \hphantom{000}1/(GHz/s) & \hphantom{0}10/(GHz/s)  \\
Memory Depth  & \hphantom{0}1.835/Mpts & \hphantom{00}1/Mpts  \\
Dead time  & 0/{$\mu$}s & \textless6/{$\mu$}s\hphantom{0}  \\
Band Width  & 500/MHz & 1/GHz\hphantom{0} \\
Sensitivity  & 1/(mV/div) & 2/(mV/div)\hphantom{0} \\
Weight  & 680/gr & 18/Kg\hphantom{0} \\
Price  & 50/K & 150/K\hphantom{0} \\
\bottomrule
\end{tabular*}
\end{center}

\section{Results and Discussions}

%

A detailed measurement on after pulse time and charge pattern of a 8
inch hemispherical PMT R5912\cite{lab11} was tested with this
measurement system.Not only the characteristic of after pulse, but
also other information could be acquired by this test system, such
as the signal photoelectron spectrum(SPE), TTS, dark current and so
on.

\subsection{Timing and Charge Distribution of after pulse}

Fig.~\ref{fig2} shown the timing pattern of after pulses of R5912.
The time of the after pulse was defined as the time interval between
the main pulse and after pulse, time difference larger than 500 ns
was shown in the plot. There were two groups which have distinct
characteristics, representing two different types of ions. For
R5912, two ion peaks were at around 1.56 $\mu$s and 6.37 $\mu$s,
corresponding to methane and caesium ionization\cite{lab12,lab13}.
This results were consistent with oscilloscope test
results\cite{lab14}. A scattered plot of after pulse charge in units
of p.e. versus arrival time was also shown in Fig.~\ref{fig2}. Most
after pulses are single p.e. charge pulses, irrelevant to main pulse
charge. The temporal uniformly distributed single p.e. charge hits
were caused by PMT dark noise.
\end{multicols}
\begin{center}
\includegraphics[width=15cm]{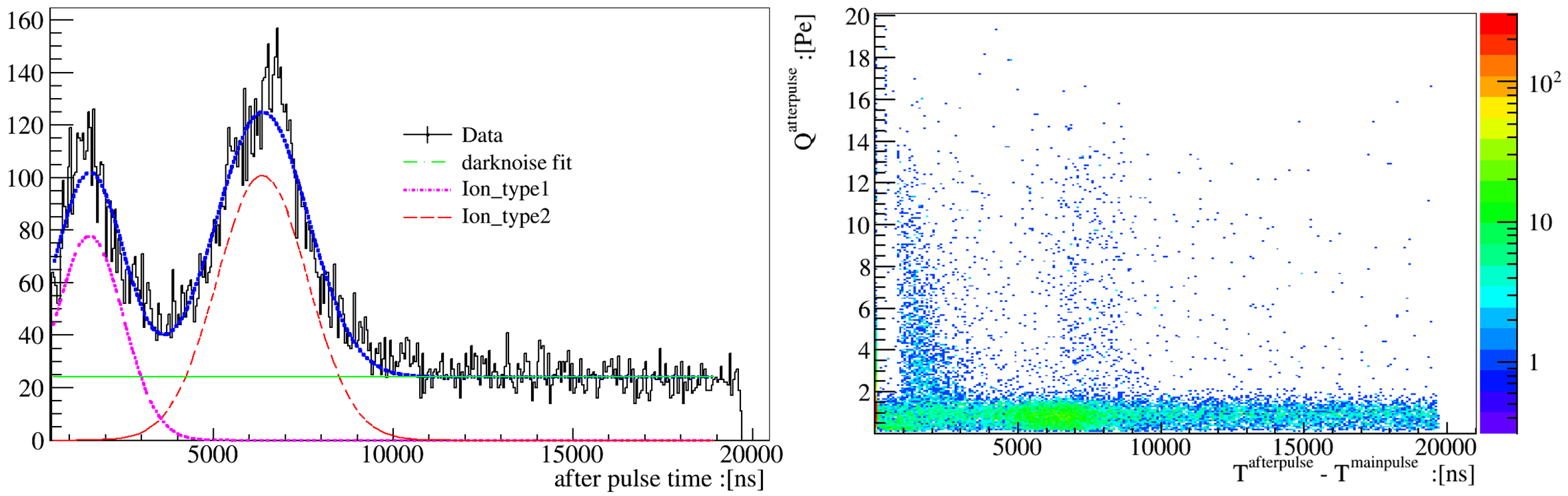}
\figcaption{\label{fig2}  (left) After pulse time distribution.
(right) After pulse time(x axis) and charge (y axis in units of
p.e.) distribution. The long and narrow band was dark noise. }
\end{center}

\begin{multicols}{2}

Besides the after pulse signal reported in
Ref.~\cite{lab12,lab13,lab14} , we had also observed a fast
component after pulse, which occurred at around 50 ns after the
initial pulse, mainly single p.e. pulses. Since many PMT test
systems use discriminator, this type of after pulse can be
suppressed completely by the discriminator dead time ($ \approx $200
ns)\cite{lab15}. This type of after pulses' charge and time
characteristics were presented in Fig.~\ref{fig3} .

\end{multicols}
\begin{center}
\includegraphics[width=13.5cm]{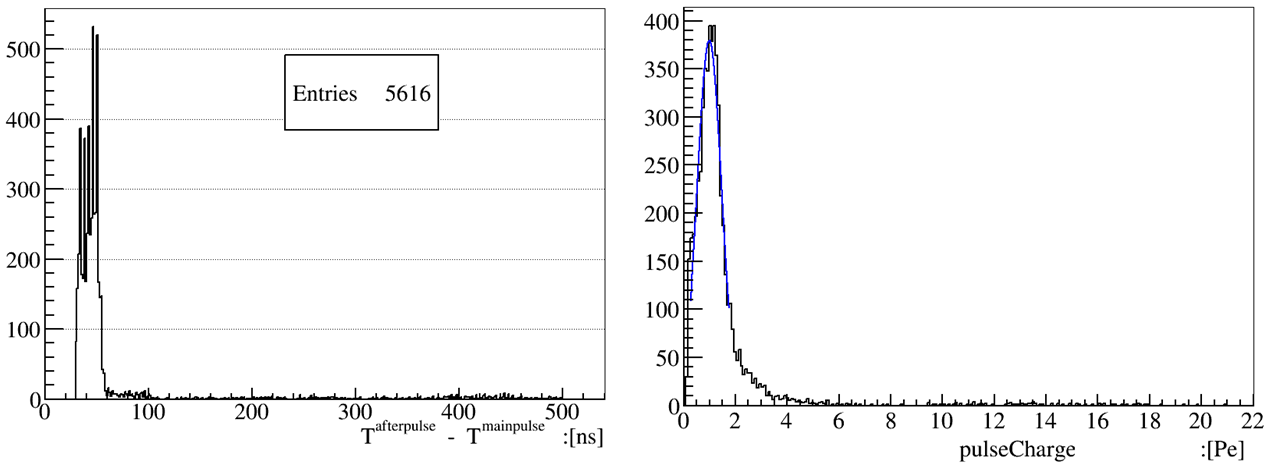}
\figcaption{\label{fig3}  (left) Fast after pulse time distribution
(right) Fast after pulse charge distribution in units of p.e. }
\end{center}

\begin{multicols}{2}

\subsection{ Absolute transit time and transit time spread}
The fast component after pulse arised from photoelectrons hitting
the first dynode backscattered. The backscattered photoelectrons
were decelerated by the electric field and then accelerated again
towards the first dynode\cite{lab16}. Thus the resulting delay time
were twice the transition time of photoelectron from the
photocathode to the first dynode. As a result, absolute transition
time and transition time spread can be obtained from hit time
distribution of fast after pulse. The results were shown in
Fig.~\ref{fig4} and Table~\ref{tab2}. Using the results from high
statistics after pulse, the absolute transit time for R5912 was 22.0
ns and transit time spread is 2.8 ns.

\begin{center}
\tabcaption{ \label{tab2} Fitting results of hit time using three Gaussian functions.}
\footnotesize
\begin{tabular*}{80mm}{c@{\extracolsep{\fill}}ccc}
\toprule Item\hphantom{00000} & Main peak/ns & AP peak /ns & 2nd AP peak/ns  \\
\hline
Hittime\hphantom{00} & 295.7 & 339.7 & 387.8\\
Spread\hphantom{000} & 5.6 & 8.6 & 10.7\\
\bottomrule
\end{tabular*}
\end{center}

\begin{center}
\includegraphics[width=7cm]{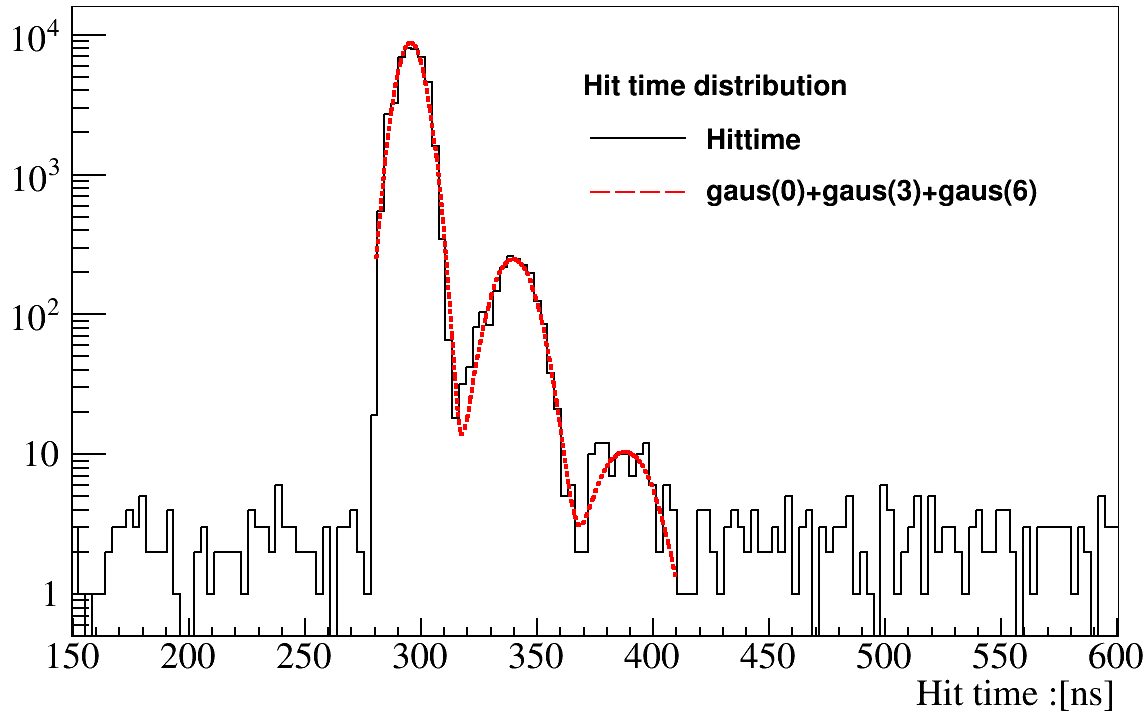}
\figcaption{\label{fig4}    Hit time of three peaks: main
pulse,after pulse, after pulse caused by after pulse. The dashed
line was fitting curve using three Gaussian functions.}
\end{center}

\subsection{ After pulse rate and dark rate}

There were various after pulse rate definitions in previous
studies,such as defining the total after pulse charge and main pulse
charge ratio as after pulse rate, or regarding after pulse number
and main pulse charge ratio as after pulse rate\cite{lab17}. Here we
defined the after pulse rate as the number of after pulse seen by
per p.e. of the initial pulse, same definition as in
Ref.~\cite{lab14}, after pulse rate was proportional to main pulse
charge, see in Fig.~\ref{fig5}. By this definition, we derived the
after pulse rate from a one degree polynomial fit to the main pulse
charge versus after pulse number plot. The offset of the linear fit
function can
 be regarded as dark count, also illustrated in Fig.~\ref{fig2}, which is
 independent of the main pulse charge. The slope of the linear fit
 function was after pulse rate, i.e. 1.794\%. This result was in consistence
with Ref.~\cite{lab14}, 1.7\%. From the offset and inspection time
period, we can get PMT dark rate. In our test, the time window was
20 $\mu$s, thus PMT dark rate at this temperature was
0.235/($20\times10^{-6}$), i.e. 11.75 KHz.

\begin{center}
\includegraphics[width=7cm]{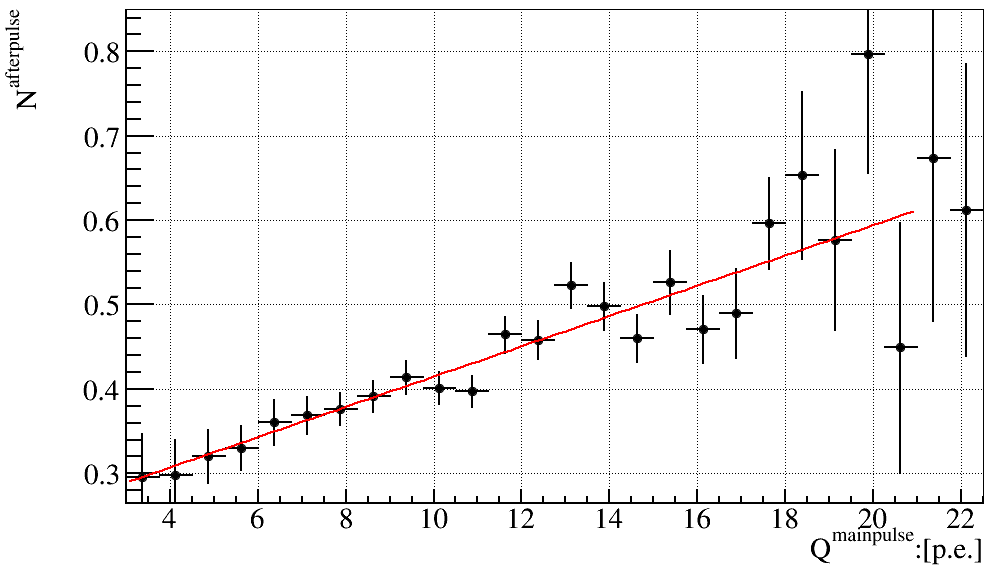}
\figcaption{\label{fig5}    After pulse number as the function of
main pulse charge in p.e.}
\end{center}

\subsection {Single photoelectron spectrum and waveform}
Typical single photoelectron spectrum system\cite{lab18} was set up
as in Ref.~\cite{lab18}. High precision electronics can be obtained
by this system. However employing NIM crate and VME framework made
the system lack of flexibility and mobility. Using the light
weighted FADC system, PMTs can be tested almost anywhere and
anytime. In the test procedure, we measured the single photoelectron
response to calibrate the FADC. To do this, we turn our light source
intensity down to ensure that SPE pulses' occupancy was less than
10\% among all triggers.

 Single photoelectron (SPE) study gave the
distribution of the charge gain and pulse peak value.From averaged
SPE waveform (Fig.~\ref{fig6}), pulse peak and pulse rise time can
also be measured, which serve as references for the PMT readout
system design.
    To detect peaks, a scan for threshold crossing in pedestal-subtracted waveform was performed.
 Some algorithm was applied to extract peaks from the waveform and get charge and hit time for each peak.
 From the averaged single p.e. waveform, we could derive some information, such as the rising time was 4.8 ns
 and the SPE amplitude is 10.897 mV, etc.

 A similar charge calculation algorithm bearing a resemblance
  to V965 QDC, produce a single photoelectron charge spectrum. From the spectrum, the common cited
  Peak/valley ratio was 3.821. And PMT charge gain can be calculated from the charge spectrum
   by doing a double Gaussian fit to the spectrum:

\begin{eqnarray}
\label{eq1} Gain &=& \frac{Q^{spe}-Q^{pedestal}}{Q^{electron}}.
\end{eqnarray}

Here, Q$^{spe}$ and Q$^{pedestal}$ was from double Gaussian fit, the
gain by DT5751 is 6.375 $\times 10^6$, with agreement to the result
from V965 6.438 $\times 10^6$ , shown in Fig.~\ref{fig7}.
\begin{center}
\includegraphics[width=7cm]{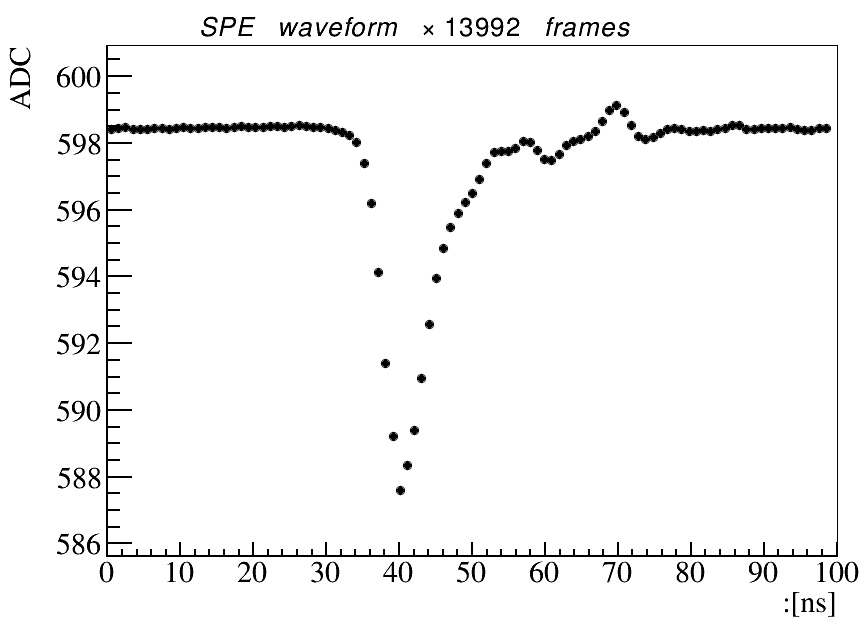}
\figcaption{\label{fig6}   Typical SPE waveform( averaged from 13882
frames ).}
\end{center}

\end{multicols}
\begin{center}
\includegraphics[width=14cm]{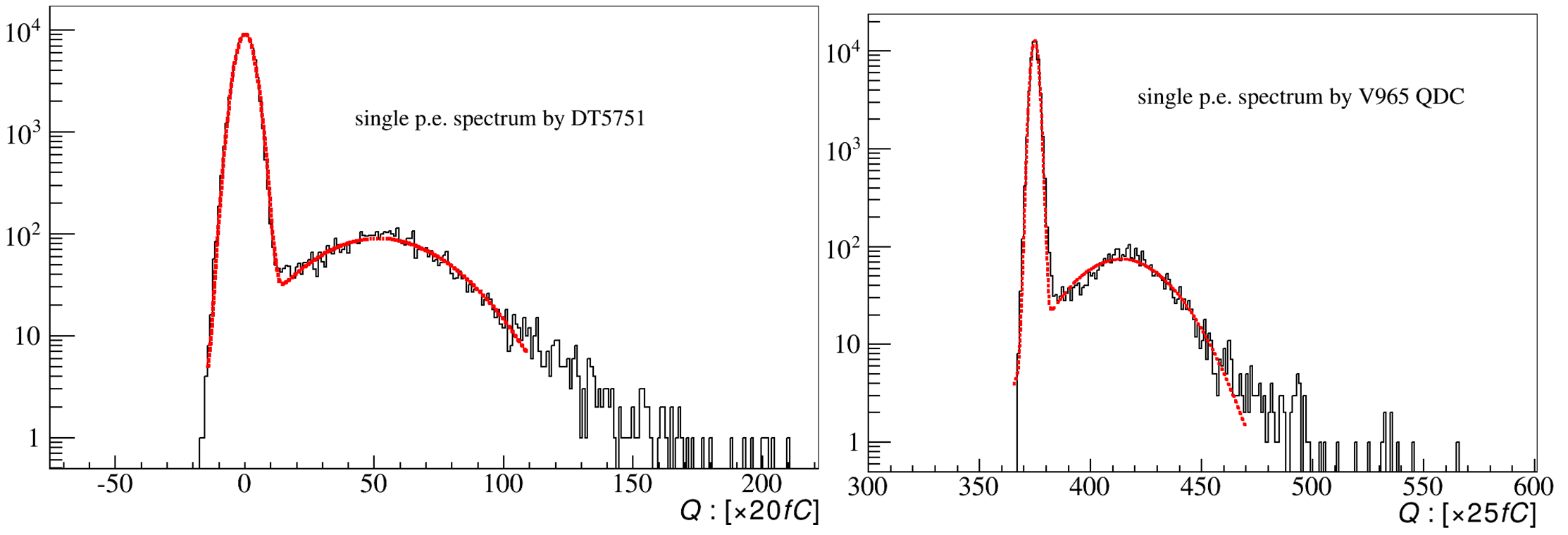}
\figcaption{\label{fig7} (left) charge spectrum of SPE by DT5751
(right) charge Spectrum by V965 QDC.}
\end{center}

\begin{multicols}{2}

\section{Conclusion}

We studied the possibility to measure after pulse with FADC and
design an experiment scheme which can finalize after pulse test in
almost one hour taking advantage of zero dead time of Flash ADC.
Meanwhile, it took over ten hours for oscilloscope to get equivalent
statistics of after pulse. We measured one 8 inch Hamamatsu PMTs'
after pulse with high precision and get plenty of useful information
by employing waveform analysis. These characters of every single
PMT, such as gain, dark rate, pre-pulse and after pulse, absolute
PMT transition time and its spread can be derived by applying this
FADC test system. The system can help PMT performance study and will
play an important role in batch PMT testing for next generation
reactor neutrino experiments, such as Jiangmen Underground Neutrino
Observatory. Since waveform sampling is the trend for high energy
physics experiment read-out system, such studies based on standalone
FADC is a salutary attempt.

\section{Acknowledgments}

The project supported by theNational Natural Science Foundation of
China (Grant No. 10775181), and the Strategic Priority Research
Program of the Chinese Academy of Sciences ( Grant No. XDA10010200
\& No. XDA10010400).

\end{multicols}
\vspace{-1mm}
\centerline{\rule{180mm}{0.1pt}}
\vspace{2mm}

\begin{multicols}{2}

\end{multicols}

\clearpage

\end{document}